\documentclass[12pt]{article}

\def\b{\begin{equation}}
\def\e{\end{equation}}
\begin{document}
\title{Comments on the proposal of Dark Energy Stars by Chapline} 
\author
{Abhas Mitra
\\
\normalsize{Nuclear Research Laboratory}\\
\normalsize{Bhabha Atomic Research Center, Mumbai- 400085, India}\\
Email: amitra@apsara.barc.ernet.in
}
\date{}

\maketitle

\begin{abstract}
We point out that the contentions of Chapline (astro-ph/0503200)[1]
that (i) General
Relativity (GR) is in conflict with Quantum Mechanics (QM) 
and (ii) GR fails even on macroscopic length scales are incorrect
and  based on (i) misunderstanding of GR and (ii) completely
misconceived ``thought experiments''. Even without introduction
of any QM, GR actually does not allow formation of any
 Event Horizon and is competent to figure out the actual nature
of the supposed Black Hole candidates. The resultant picture is not
in conflict with QM at all. 
 \end{abstract}


\section{Introduction}
In a recent preprint (astro-ph/0503200) Chapline[1] has proposed
that the supposed Black Holes(BHs) are actually compact stars
filled with dark energy and having a physical surface unlike BHs.
 He has termed such compact objects as ``Dark Energy Stars''.
Actually it was already shown several years ago that the 
supposed BH candidates cannot be true BHs and
must be Ultra Compact Objects (UCOs) with physical surface[2,3,4,5]. Here
Chapline has insisted for a specfic QM model of such compact objects
on the plea that GR is in conflict with QM and hence must be invalid.
We systemetically show below that this contention of Chapline is completely
unfounded and his specfic line  of  thinking behind 
``Dark Energy Stars'' is based on elementary but serious errors.
This is not to tell that  there cannot be any Quantum Gravity (QG)
inspired model of UCOs.

\section{Synchronous Time}
Chapline insists  that ordinary QM requires an universal  time
which is present in Minkowski spacetime.
While it is obvious that 
 the non local large scale correlations implied by QM could be easily explained in
terms of an univeral  time,  this, however, does not mean that in the absence
of an universal time, such correlations would cease to exist. On the
other hand, it would  just be more difficult to formulate and work out those
correlations.

In fact even if there would be no gravity and no GR, and  electrons
or anthing else would have accelerated motion  with respect to each other, the
clocks attached to  them would not be universally synchronized. Will it
mean that QM is in conflict with 
the occurrece of accelerated motion? Will it mean that there cannot
be any accelerated motion?

A logical extension of Chapline's concern would be that, in a strict sense,
 in the presence of gravity, one needs to have an appropiate theory of Quantum Gravity 
(QG).
But coming back to Chapline's immediate concern, effect of varing gravity
could actually be not much significant for long range QM correlations. Further,
in the context of Friedmann model, the cosmic time may actually serve as a univeral
time (if one would ignore lack of clock synchronization due to the
far more serious proper mutual accelerations).

  To show that GR may not always admit universal clock synchronization, Chapline,
rather unnecessarily mentions of Godel's universe. Chapline argues that since Godel's universe allows closed time like curves
GR must be erroneous. Godel's universe is filled with pure ``dust''
having pressure $p\equiv 0$[6].
 If ``dust'' is to be treated as a fluid, as is the requirement for Einstein (
hydrodynamic)
equations, thermodynamics would tell that one can strictly
have $p=0$ only if the energy density of the fluid $\rho =0$. In other words Godel's universe
is only a mathematical model which corresponds to no mass energy at all. When there is
no mass energy, there is no rotation, no physical universe either!
Thus it is not GR which allowed closed time like curves; on the
other hand, it is the incorrect assumption of a dust model ($p\equiv 0$)
for a   phycial cosmic fluid
 which created this incorrect impression.

Chapline correctly points out that whenever rotation is involved,
universal clock synchronization would not be allowed in GR. But this will be true
for not only GR but probably for all relativistic theories of gravity.
In fact what Chapline appears to be unaware of is that even if one would ignore
GR or any other relativistic theory of gravity and consider only
Special Theory of Relativity, the spacetime cross term in the metric cannot be
done away with and {\em there would be no universal time}[7]. So, then,
will Chapline demand that even special theory of relativity
 be abandoned?

And Chapline's comment that GR is not valid in cosmology context
(too) is simply unsubstantiated.

\section{Spacetime Superfluidity?}
Here Chapline starts with the assumption that ordinary matter be viewed
as excited states of vacuum. Why? Quantum vacuum fluctuations  may give
rise to various virtual pairs and those pairs may be considered
excited states of vacuum. But why normal matter? What about
baryon and lepton number conservation in QM? In fact, here, Chapline
seems to overlook conservation principles of QM altogether.

Following this, Chapline proposes that all atomic motions must be
irrotational. But we know that it is possible to have eddies and vorticities in
air, water and all ordinary fluids  where $\nabla\times {\vec v} \neq 0$.
Most likely Chapline is inspired here by a footnote of Landau \& Lifshitz[8]
which points out the simple fact that if the fluid is moving in such a manner
that it is possible to have a global synchronization of clocks
(i.e., if the spacetime cross term can be eliminated, which is possible
when there is no rotation), the fluid motion will be irrotational:$\nabla \times {\vec v} = 0$.
Actually, this fluid motion {\em does not at all refer to any mysterious vacuum motion}.
But Chapline has attributed this natural and obvious result to the motion
of the vacuum itself, when, actually we are concerned here the motion
{\em of a fluid in the spacetime}.
Clearly, Chapline's  formulation of the model has hardly any physical basis.
Even if one, momentarily, admits that all motions in the universe is
irrotational, Chapline's conclusion that, hence ``an appropriate model for the
vacuum of spacetime is a superfluid'' is too far fetched at the best. In any case,
this result is obtained by confusing the motion of the fluid with the motion of
the vacuum! 

And even if we accept this fluid motion (equated with vacuum motion) is irrotational
why should vacuum  be considered as a superfluid?
If such an equivalence would exist, then {\em any superfluid 
may be considered as vacuum as well}!
Further, in the first place, why not any irrotational  fluid 
 (which need not be a superfluid at all)
be considered as a superfluid? Thus there is no physical basis for such a
connection between vacuum and superfluidity.

\section{Chapline's Thought Experiment}
Chapline purports to do a thought experiment to show ``why it is
wrong to assume that classical GR is always correct on macroscopic length
scales'' He imagines a vertical column of Bose superfluid; then he conceives
that ``As a result of the increasing pressure in the fluid as a function
of depth it could happen that at a certain depth the speed of sound vanishes''!
How? With increasing depth, density can only increase and elementary physics
tells that sound speed will {\em increase rather than decrease}. It becomes clear
subsequently, why, despite this, Chapline wished the sound speed to decrease
and that too to zero value: because he thinks that speed of light becomes zero
at the EH of a BH described by the metric:

\b
ds^2 = -\left(1 - {\alpha_0\over R}\right) dt^2 
+\left(1- {\alpha_0\over R}\right)^{-1} dR^2 + R^2(d\theta^2 + \sin^2\phi d\theta^2)
;~~ R \ge R_0=0
\e
Here, $\theta$ and $\phi$ are the polar angles and the {\em integration constant}
$\alpha_0$ is interpreted as twice the gravitational mass of the ``Massenpunkt'':
 $\alpha_0 
= 2M_0$ ($G=c=1$). The metric coefficients are

\b
g_{TT} = -\left(1 - {\alpha_0\over R}\right); \qquad g_{RR} = \left( 1- {\alpha_0\over R}\right)^{-1}
\e

It follows that as a test particle moves in the spacetime described by this metric its
 {\em coordinate speed}, $dR/dT =0$ at the EH, not only for photons but for anything.
On the other hand, the physical 3-speed:
\b
v = {\sqrt{g_{RR}}~dR \over \sqrt{-g_{TT}} ~dT} =1
\e
for photons everwhere including the EH, {\em by definition}. Incidentally, the 3-speed
of any material particle too acquires the speed of light at the EH.

In a complete misunderstanding of GR
Chapline thought that the physical speed $v=0$ for photons at the EH and also 
wished $c_s =0$ at the same
place to further build up his idea that vacuum behaves like a superfluid with
strange critical properies.
In reality, neither  $v$ nor $c_s$ is zero  at the EH. Also, even if they were zero, there would not
be any physical basis for connecting the two different phenomenon.

Now from all angles, we have found that  the inevitability
of ``Dark Energy Stars'', as discussed by Chapline, is based on elementary but severe 
misunderstanding and misconception. Thus there
is no basis for thinking that GR is not valid even on macroscopic scales,
vacuum behaves like a superfluid and there could be dramatic QM effects
at large $R$.

\section{Chapline's Dark Energy Star}
Apart from the erroneous formulation of the concept behind ``Dark Energy Stars'',
there is a general inconsistency in the whole scheme:
While Chapline starts with the premises that ``EHs cannot exist in real world'', 
he eventually gives the sketch of a model of a star whose boundary is nothing but
the EH! Chapline simultaneously rejects and accepts the EH. The latter necessity arises
because he wants the boundary of the proposed ``Dark Energy Star'' to be the QM critical
surface where time dilation factor approaches zero. Chapline mentions that time dilation
factor ($\sqrt{-g_{TT}} $) inside the EH of a BH is ``negative'' whereas it is actually
imaginary (for the metric [1]).

Probably Chapline is unaware of the fact that the entire region beneath the EH is a
{\em trapped region}. And hence his dark fluid may not stay put and keep filling in
the interior region. In a trapped region, for all realistic equation of state, all
mass energy must undergo  collapse inexorably  to the central singularity[9]. Thus within
no time, his Dark Energy Star may reduce to a BH, if there would be an EH at a finite $R$!

\section{GR or Event Horizon to be blamed?}
There is however one apparently valid basis for some of the conjectures made by
Chapline. He notes that, although, it is asserted that the EH is a mere coordinate
singularity and, physics-wise, nothing unusual happens there, actually, lot of
unusual physical effects take place at the EH. Chapline mentions that QM Green's function
develpos a ``cusp-like behaviour'' at the EH, where one can make $K_{EH}$ arbitrarily
small {\bf if $M$ were really arbitrary parameter} for a BH.

From this Chapline concludes that ``it is wrong to assume that classical GR is
always correct on macroscopic length scales''.

This also leads him to ``question whether quantum corrections to classical
GR can be important under circumstances where at least locally spacetime appears
to be quite ordinary''.

First let us see that unusual physical results appear at the EH not only from
the point of view of QM. For instance, as noted by Chapline, the covariantly defined
quantity surface gravitational redshift (which for Earth,  is $z_E \approx 0$)
\b
z_{EH} = \infty
\e
for a BH.  $z$  can be measured by any distant observer and is very much a physical
quantity. Then why does it diverge at EH? And this is definitely not due to any QM effect.

From the acceleration 4-vector of a test particle, one can construct an
{\em Acceleration Scalar}:

\b
a= {M\over R^2\sqrt{1- 2M/R}}
\e

On the EH, this {\bf scalar diverges}, $a_{EH} =\infty$[4,10]! Neither is this due to
any
 QM effect. To see more of it consider a dilute non-interacting gas
of dimension $R_{max}$
 around an object of circumference radius $R_0$. Let $m$ be the rest mass and
$E$ be the constant energy of the molecules of the gas.

The {\em phase volume} of this gas, an {\em INVARIANT}, is given by[11]:

\b
\Gamma(E) = {16 \pi^2\over 3} \int_{R_0}^{R_{max}} {R^2 ~dR\over \sqrt{1- 2M/R}}
\left({E^2\over 1 -2M/R} - m^2\right)^{3/2}
\e
The derivative of $\Gamma(E)$ w.r.t. $E$ gives the entropy, $g(E)$ of this gas,
and which is another {\em INVARIANT} for the system.

As long as $R_0 > 2M$, i.e., if the object is not a BH, then $\Gamma(E)$ and
$g(E)$ behave nicely. But if this object would be assumed to be a BH with finite
$M$, i.e., if $R_0 = 2M >0$, then, both these {\em scalars} would {\bf diverge}[11].

So, if it would be assumed that there is a BH of arbitrary finite mass $M$,
then any number of unacceptable physical effects would happen even at a pure
classical level. Thus occurrence of such unphysical behaviours cannot be
attributed to any QM phase transitions.

On the other hand, it is obvious that, it is the assumption of occurrence of EH at 
arbitrary
finite value of $R$, which is responsible for such occurrences. At this stage
one would argue that  by virtue of the {\em completely} vacuum Schwarzschild solution (1) GR
gives rise to arbitrary finite value of $R_g = 2M$. Therefore, ultimately, it is
GR which is responsible for these unphysical behaviours. And this would be cited
as a proof for ``failure of classical GR on macroscopic length scales''.
The basic problem lies at a more fundamental classical level. 
Let us analyze the scenario more closely.

GR says that the best measure for the strength of gravitation is
the value of the Kretschmann {\em Scalar}, $K$. For a static spherical object of circumference
radius $R_0$, we have

\b
K = {48 M^2\over R^6}; ~\qquad R \ge R_0
\e

Let on the surface of Earth, $K=K_E$. In case the object is a Schwarzschild BH, the interior region too would be vacuum
and the foregoing relation would be valid at any $R$. In particular, at the EH, $R=2M$,
we have
\b
K_{EH} = {3\over M^4}
\e

 Chapline correctly points out that for a sufficiently massive BH, $K_{EH}$ could be
very small. Let
\b
K_{EH} = b K_E
\e
where $b$ could be {\em arbitrarily small}, i.e., {\em gravity on the EH or in the interior region of a BH}
 could be {\bf infinitely
weaker than on the surface of Earth} and  yet one can have

$\bullet$ (i) covariantly defined and physically measurable $ z= \infty$, 

$\bullet$ (ii) Phyically
measurable {\bf scalar} $a=\infty$, 

$\bullet$ (iii) physically measurable {\bf scalar}
$g(E) =\infty$ and as mentioned by Chapline 

$\bullet$ (iv) QM Green function blowing up too.

Let us now recall  the popular notion  of a BH/EH:

``...{\em where gravity is so strong}(!) that even light cannot escape..''.

One should wonder then how can a gravity which {\em could be infinitely weaker than
the gravity on Earth's surface} be ``{\em so strong} and  do all such tricks (i-iv). And if such
unusual things can indeed happen in arbitrarily weak gravity, then

``Why our Earth with an infinitely stronger gravity would not be able to trap photons
 permanently
and why should we not  have $z=a=g =\infty$ on Earth's, Moon's or Sun's  surface too?''

But we know that no such spectacular things happen in the weak gravitational field
of Earth: GR really does not do such odd things or fail at large $R$.

Hence the big question here ought to be ``Does GR Really Allow  Existence 
of Finite Radius EHs or Finite Mass BHs?''

The very fact that the scalar $a$  blows up at the EH immediately signals
the fact that the EH is not region of mere ``coordinate singularity'', on the
other hand, it must be the true singularity! But the true singularity is at
$R=0$, Thus one must have $R_g =0$, or, $M=0$. And when this is so, it can be found
that
\b
K_{EH} \equiv \infty
\e
as well as
\b
a= \lim_{R\to 2M \to 0}{1\over 4M \sqrt{1 -2M/R}} =\infty
\e
And this is the reason why all those unusual physical things were found to be happening
at the EH! Gravity is not arbitrarily weak there, but, on the other hand, gravity
is {\em indeed infinitely strong there}!  Therefore,  in a consistent manner, at the EH,

${\bullet}$ The Lorentz Factor of an Incident Particle, $\gamma =\infty$

$\bullet$ The Local Energy of an Incident Particle, $\cal E =\infty$

$\bullet$ The Spectral Lines suffer a Redshift of $z=\infty$

$\bullet$ Acceleration Scalar $a=\infty$

And most importantly, in complete consistency,

$\bullet$ Kretschmann Scalar $K=\infty$

$\bullet$ Only when,  Mass of  BHs $M =M_0 \equiv 0$

Thus GR did not fail when it allowed those unusual
physical events at the EH.

Incidentally, when the BH mass $M=0$, the the {\em invariant} phase space volume of
the surrounding gas becomes non-singular:

\b
\Gamma(E) = {16 \pi^2\over 3} (E^2 - m^2)^{3/2}
\int_0^{R_{max}} R^2 ~dR ={16 \pi^2\over 9} (E^2 - m^2)^{3/2} {R_{max}}^3
\e
Accordingly, the {\em invariant} entropy of the gas too would become finite only
if $M=0$ (note that, if $R > 2M$, then neither $\Gamma$ nor $g$ vanish for finite
$M$). Also bear in mind that $\Gamma$ and $g$ are not the scalars associated with
the EH and hence they must not blow up. On the other hand, they are properties of
the surrounding gas and must be finite (for a finite $R_{max}$) provided
the underlying physics problem is formulated consistently, i.e., when, $R_g =0$.

Since much of the impetus for Chapline's model comes from the apparent idea
that GR predicts existence of finite mass BHs and finite radius EHs, it would be
worthwhile to clinically show that the value of the integration constant occurring
in the completely vacuum Schwarzschild Soln. (i.e, a BH) is $\alpha_0\equiv 0$.

\section{Invariance of 4-volume}

The original identification of $\alpha_0= 2M$ was done by matching the vacuum
solution in Metric (1) with the corresponding Newtonian solution at large $R$.

 Newtonian gravitation allows for the existence of a
spherical ``point mass'', i.e, $R_0 =0$, and since  in Newtonian gravitation, mass is essentially
of baryonic or leptonic origin (bare mass) with no negative ``dressing'' due to gravity or any self-energy, 
all masses including that of a ``point'' is necessarily finite, $M_0 > 0$.
And, in GR too,  it is has so far been {\bf assumed} that $M_0$ would 
continue to be finite
even when the spacetime is completely empty, i.e, $R_0 =0$ (mass point). 
It is this expectation which
gave rise to the  concept of Black Holes in the GR era[12].

The determinant of the Schwarzschild metric (1) can easily be found to be
\b
g_1 =-R^4 \sin^2 \theta 
\e

When the Schwarzschild solution represents pure vacuum, both in the {\em interior} as well as in the exterior, 
then, by integrating,
the {\em vacuum} photon propagation equation ($dR/dT$), from $R=R_0=0$ to $R=R$ (a procedure
which would not be valid if the Schwarzschild solution would represent only the vacuum
spacetime outside a spherical object filled with mass energy, i.e., if $R_0 > 2M =finite$)
one obtains various special coordinates to study the BH  spacetime. One such coordinate
system is the socalled Eddington-Finkelstein coordinates: 

\b
t_* = t \mp \alpha_0 \log \left( {R\over \alpha_0} -1\right)
\e
\b
R_* = R;\qquad \theta_* =\theta;\qquad \phi_* =\phi
\e

In terms of these coordinates, the BH metric becomes
\b
ds^2 = - \left(1 - {\alpha_0 \over R}\right)dt_*^2 \mp {2\alpha_0\over R} dt_* dR 
+ \left(1 +{\alpha_0\over R}\right)dR^2 + R^2(d\theta^2 + \sin^2\phi d\theta^2)
\e

The corresponding metric coefficients are
\b
g_{{t_*} t_*} = -(1-\alpha_0/R), ~~ g_{RR}=(1+\alpha_0/R),~~ g_{{t_*} R}=g_{R t_*} 
= \pm \alpha_0/R
\e 
In this case the determinant is same as $g_1$:

\b
g_* = -g_{\theta \theta} g_{\phi \phi}(g^2_{t_* R} -g_{t_* t_*}
 g_{RR})
= -R^4 \sin^2 \theta = g_1
\e
 Now let us apply the principle of {\em invariance of 4-volume}[7,9] for the coordinate
systems ($t$, $R$, $\theta$, $\phi$) and ($t_*$, $R$, $\theta$, $\phi$) 
 at arbitrary $R$ and {\bf not necessarily at $R \le 2M$}:
\b
\int \sqrt{-g_4}~ dt_*~ dR~ d\theta ~d\phi = \int \sqrt{-g_1}~  dt~ dR~ d\theta~ d\phi
\e
Since $g_1= g_2 = -R^4 \sin^2 \theta $, after doing the
$\theta$ and $\phi$ integrations
 we promptly obtain,

 \b
\int R^2 ~dR ~dt_* = \int R^2~ dR~dt
\e

From Eq.(14), we also have

\b
dt_* = dt \mp {\alpha_0 \over R - \alpha_0} dR 
\e

By substitution of this foregoing simple relation in Eq.(20) we find
\b
  \alpha_0\int {R^2~dR \over R - \alpha_0}  = 0
\e

And this shows that
\b
\alpha_0 \equiv 0
\e
(Remember again that, if there would be mass energy upto $R =R_0 >2M$, i.e., for the 
case which involves mass-energy, this result
would not be valid).

Thus, {\em in a most direct manner}, we obtain the fundamental result that the mass of
the Schwarzschild BHs:
\b
M_0 = M(R_0=0) = \alpha_0/2 \equiv 0
\e

\section{Implications}
Since all BH candidates or any other objects in the universe have $M >0$, they
cannot be BHs. Thus most of the concerns of Chapline actually get solved by
properly interpreting GR and there is no need to belittle or suspect GR.

With $\alpha_0 \equiv 0$,   the EH merges with the central singularity at $R=0$ and
hence the metric (1) has only one singularity, the 
singularity  at $R=\alpha= \alpha_0=0$. This fundamental result is not at all in
conflict with QG because many Supersymmetric String Theories obtain copious
solutions for Extremal BHs where {\em the horizon merges with the central singularity}[13]
as found here classically!

 We may recall that there is only one {\em exact analytical} solution of spherical
gravitational collapse where an uniform dust of mass $M$, initially ($t=0$) 
at rest with a radius $R_i$, collapses to a SBH in a proper time[14]
\b
\tau_c = {\pi\over 2} \left({R_i^3\over 2 GM}\right)^{1/2}
\e
Since the dust ball is at rest at $t=0$, we can use the equation for hyrdo-static
balance[15] at $t=0$:
\b
{dp\over dR} = - {p +\rho_i\over R(R-2M)} (4\pi R^3 p + M)
\e
 For a dust $p\equiv 0$, and therefore,
 we have $dp/dR \equiv 0$, and then the foregoing Eq. yields
$\rho_i =0$. From thermodynamical point of view too, whether at rest or not,
a $p=0$ equation of state is physically obtained only if $\rho =0$. Therefore,
trivially, the mass of the dust ball is zero for a finite $R_i$. Hence the mass of the
resultant SBH is indeed $M=0$. 
But this was always ignored, it was pretended that eventhough, $p=0$, we must have
$\rho =finite$ and $M$ too would be finite in tune with the idea that the integration
constant $\alpha_0 >0$ (when it was actually zero).

And when $M=0$, from Eq.(25), it follows that,
$\tau_c =\infty$. Thus though the $M=0$ SBHs are, mathematically, allowed
by GR, they  cannot be realized in an universe with finite proper age.
 
Since there is no EH in a finite proper time, there is no trapping or loss of
quantum information either and hence another worry of Chapline also gets
solved in a purely classical manner.

It was shown earlier that even for the most general case of spherical
collapse (i.e, not necessarily for uniform dust), no trapped surface is ever formed[3,4,5,16].
Thus a collapsing fluid must always radiate and if it would be assumed to undergo
continued collapse , we would have 
$M\to 0$ asymptotically. This is the reason  that the integration
constant $\alpha_0 = 2M_0$ turned out be identically zero. Hence the observed BH candidates with masses often much higher
than the upper mass limit of {\em cold baryonic} bodies in hydrostatic equilibrium,
cannot be SCBs. Detailed analysis of recent observational data indeed suggests that
the BH candidates have strong intrinsic magnetic moments rather than any EH[17,18,19].
 And it
has been suggested that the BH candidates are Magnetized Eternally Collapsing Objects
(MECOs). These are extremely {\em hot} objects in
quasistatic equilibrium due to
extremely strong radiation and magnetic pressure[17,18,19]. However, if Quantum Gravity
would be invoked, the supposed Black Holes could be {\em cold} and {\em static} configurations
with hard surfaces[20] where this boundary hard surface must not be any EH.

 Chapline firmly thinks that GR should not be taken seriously ``on any macroscopic length
scale'' and by seeing this not to be the case, he writes

``Indeed I am sure it will be a puzzle to future historians of science as to why it
took so long to realize this''!

Future historians of science might also wonder when so much literature showing that
GR actually does not allow formation of EH (at a finite $R$) was already available
(as shown in the bibliography of this preprint),
how Chapline was unaware about the same.




\end{document}